\documentclass[aps,10pt,pra,notitlepage,twocolumn,noeprint]{revtex4-1}
\usepackage{dcolumn}
\usepackage{graphicx}
\usepackage[T1]{fontenc}
\usepackage[utf8]{inputenc}
\usepackage{multirow}
\usepackage{lipsum}
\usepackage[bookmarks=false,breaklinks,colorlinks]{hyperref}

\newcommand{\etal}{\textit{et al}.}
\newcommand{\ie}{\textit{i}.\textit{e}.}

\begin{document}
\title{Sampling dependent systematic errors in effective harmonic models}

\author{E. Metsanurk} \email{erki.metsanurk@physics.uu.se} \affiliation{Department of Physics and Astronomy, Uppsala University, Box 516, S-75120 Uppsala, Sweden} 
\author{M. Klintenberg} \affiliation{Department of Physics and Astronomy, Uppsala University, Box 516, S-75120 Uppsala, Sweden}

\date{\today}

\begin{abstract}
Effective harmonic methods allow for calculating temperature dependent 
phonon frequencies by incorporating the anharmonic contributions into 
an effective harmonic Hamiltonian. The systematic errors arising from 
such an approximation are explained theoretically and quantified by 
density functional theory based numerical simulations. Two techniques 
with different approaches for sampling the finite temperature phase 
space in order to generate the force-displacement data are compared. It is 
shown that the error in free energy obtained by using either can exceed 
that obtained from 0~K harmonic lattice dynamics analysis which 
neglects the anharmonic effects.

\end{abstract}

\maketitle

%\section{TODO}

\section{Introduction}
The harmonic approximation of the potential energy is a fundamental 
part of the analysis of vibrational properties of materials both in 
theoretical and experimental studies \cite{Fultz2010}. Its usefulness 
is multifold in the sense that not only does it allow for capturing the 
dominant part of the potential energy surface, but makes possible to 
derive exact analytical relations for all temperature dependent 
vibrational thermodynamic properties. In addition, the harmonic system 
can be used as a reference point for either integration- or 
perturbation-based methods in order to find the corrections due to 
anharmonicity that always exists in real systems.

The lattice dynamics approach \cite{Togo2015} to calculate the normal 
vibrational modes of a crystal can break down when either very strong 
quantum mechanical effects dominate as is the case for crystalline 
helium \cite{DeWette1967} or more commonly due to mechanical 
instabilities, for example cubic zirconia \cite{Parlinski1997} or 
titanium, zirconium and hafnium in BCC structure \cite{Souvatzis2008}. 
This results in some of the normal modes having imaginary vibrational 
frequencies due to decrease in the total potential energy when the 
atoms are displaced along those modes. Since the harmonic free energy 
and all the other derived thermodynamic properties are defined as 
integrals over the whole normal mode space, it would result in 
complex-valued properties and are, thus, commonly interpreted as 
undefined.

Recently, methods have been proposed to calculate free energies of such 
systems both for cases where the lattice is dynamically unstable 
\cite{VandeWalle2015} or stabilized due to temperature 
\cite{Kadkhodaei2017}. Both are based on partitioning of atomic 
configuration space, calculating the free energy of each region 
separately and adding those together consistent with equilibrium 
thermodynamics.

Instead of the aforementioned detailed mapping it is also possible to 
fit high temperature force-displacement data obtained from atomistic 
simulation to a truncated expansion of the potential energy surface 
after which the second order effective force constants can be used to 
calculate the temperature dependent phonon frequencies and, hence, the 
effective harmonic free energy. This is the basis for the methods of 
self-consistent \textit{ab-initio} lattice dynamics (SCAILD) 
\cite{Souvatzis2008,Souvatzis2008a,Souvatzis2009} and 
temperature-dependent effective potential (TDEP) 
\cite{Hellman2011,Hellman2013a,Mosyagin2016}. Neither are inherently 
limited to be used for analyzing only systems with dynamical 
instabilities, which could be considered as the most severe case, but 
any kind of anharmonicity.

Currently the best way to gather the data for either of the methods is 
through density functional theory (DFT) \cite{Kohn1965} which in 
principle provides a parameter-free way to explore the dynamics of 
realistic, as opposed to model, systems. \textit{Ab-initio} based 
thermodynamics has been shown to be a promising way to predict the 
thermal properties of materials taking all the relevant excitations, 
not limited to vibrational, into account 
\cite{Grabowski2011,Palumbo2014}.

Without considering whether DFT with its approximations can provide 
reliable enough data, there are also other systematic and statistical 
uncertainties associated with calculating thermodynamic properties. 
Some of those, such as the limited system size, can be partially dealt 
by Fourier interpolating the information about vibrations in an 
infinite crystal \cite{Parlinski1997}, while others such as limited 
amount of time for sampling can be dealt with by upsampled 
thermodynamic integration with Langevin dynamics (UP-TILD) or 
harmonically mapped averaging \cite{Moustafa2015,Moustafa2017a}. 
Regardless of the method used, the total uncertainty required to 
produce satisfactory results is often under 1 meV/atom. For example, a 
6 meV/atom shift on the energy scale results in 400~K (60\%) 
overestimation of the FCC to BCC transition temperature for calcium 
using \textit{ab-initio} calculations \cite{Grabowski2011a}.

Given the requirement of high accuracy for any method used to calculate 
thermodynamic properties it is necessary to understand how the models 
behave under certain conditions. In this study we analyze the accuracy 
of effective harmonic models by comparing SCAILD and TDEP both 
theoretically and numerically at different temperatures. It has been 
shown previously that the difference between the free energies 
predicted by the two methods increases with temperature 
\cite{Korotaev2018}, however, no theoretical explanation was provided 
what in particular could cause the discrepancy between these two very 
similar models.

\section{Methodology}
\subsection{Harmonic approximation at 0 K and finite temperatures}

\begin{figure}
    \includegraphics[width=\columnwidth]{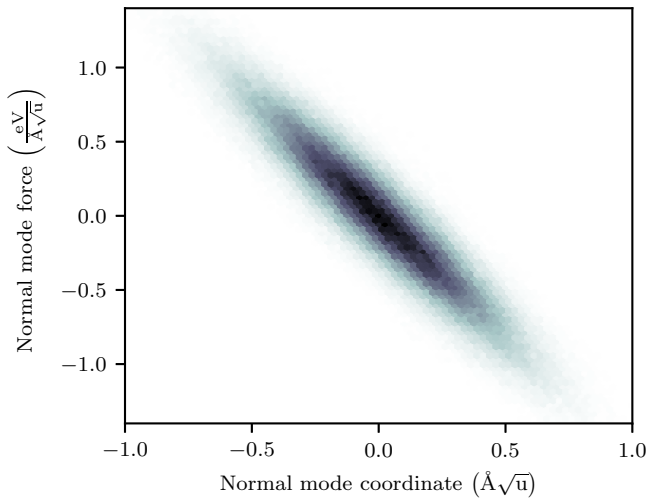}
    \caption{\label{fig:cloud_histogram} A histogram of the 
    force-displacement relation for a single phonon mode extracted from 
    a molecular dynamics simulation at 1600 K. The darker areas 
    correspond to higher probability states. A linear fit through the 
    distribution gives the negative square of the effective mode 
    frequency. }
\end{figure}

In order to simplify the comparison of the two methods a slight 
reformulation of the theory is needed compared to that described in the 
original works. The starting point is the force-displacement relation 
through a $3N \times 3N$ force constant matrix $\Phi$, so for any 
atomic configuration~$c$ the forces are given by

\begin{equation}\label{eq:forces}
\vec{f_c} = - \Phi(\vec{\theta}) \vec{u_c}
\end{equation}

In a similar fashion to Hellman \etal\ \cite{Hellman2013a}, the 
elements of $\Phi$ are expressed as linear combinations of parameters 
$\vec{\theta}$ as the force constants of crystalline systems are in general 
not independent, but constrained by rotational, 
translational and inversion symmetry of the lattice and by the 
requirement that force appears on any atom under rigid translation 
of the whole crystal. For the supercell used in this study, the number 
of force constants are reduced from $82944$ to $52$.

This allows to rewrite Equation~\ref{eq:forces} as

\begin{equation}
\vec{f_c} = C(\vec{u_c}) \vec{\theta}
\end{equation}

where the elements of matrix $C_c$ are linear combinations of 
displacements the coefficients of which depend on the choice of
the supercell and the aforementioned symmetry constraints.
The parameters $\vec{\theta}$ can now be found with linear least
squares method using Moore-Penrose inverse of $C$

\begin{equation}
\vec{\theta} = C^+(\vec{u}) \vec{f}
\end{equation}

Where we have omitted the indices $c$ since both $\vec{f}$ and
$\vec{u}$ can contain the forces and displacements of any
number of atomic configurations. The more anharmonic the system is,
the more configurations are needed in order to keep the uncertainty
of the fit sufficiently small. By expressing $\vec{\theta}$ as a function of 
temperature we obtain a temperature dependent effective harmonic force
constant matrix from which the phonon frequencies and eigenvectors
can be calculated.

%Ordinarily the phonon frequencies are calculated as a function of wave 
%vector $\vec{q}$ in the reciprocal primitive unit cell through the 
%Fourier transform of the force constants resulting in separate $3n 
%\times 3n$ dynamical matrices $D(\vec{q})$ where $n$ is the number of 
%atoms in the primitive cell. Exact phonon frequencies are calculated 
%for vectors $\vec{q}$ commensurate with the supercell by diagonalizing 
%the corresponding $D$ whereas for other wave vectors the results are 
%valid only if the supercell is large enough so the force constants 
%become zero for distant pairs of atoms \cite{Parlinski1997}.

When working only with the vibrational modes commensurate with the 
supercell it is more convenient to consider the supercell itself to be 
the unit cell and calculate the exact normal mode frequencies which 
correspond to $\vec{q} = (0,0,0)$, \ie\ only at the $\Gamma$-point of
the reciprocal lattice of the supercell. In 
that case the dynamical matrix for the whole system is

\begin{equation}\label{eq:dynmat}
D = M^{-\frac{1}{2}} \Phi(\vec{\theta}) M^{-\frac{1}{2}} = Q(\vec{\theta}) \Omega^2 (\vec{\theta}) Q(\vec{\theta})^T
\end{equation}

where $M$ is a diagonal matrix of the masses of the atoms, $\Omega^2$ a 
diagonal matrix of eigenvalues and $Q$ a matrix the columns of which 
are the eigenvectors.

By substituting Equation~\ref{eq:dynmat} into Equation~\ref{eq:forces}
we get the following general equation that applies both to SCAILD and TDEP
\begin{equation}\label{eq:frequency_fit_general}
Q(\vec{\theta})^T  M^{-\frac{1}{2}}  \vec{f_c} = - \Omega(\vec{\theta})^2 Q(\vec{\theta})^T M^{\frac{1}{2}}\vec{u_c}
\end{equation}

One of the approximations of SCAILD is that the eigenvectors are 
temperature independent and can, thus, be obtained from a quick 0~K 
phonon calculation. Whether this holds, depends on the symmetry 
properties of the system and the choice of the supercell, which 
determines the grid of sampled $q$-points \cite{Maradudin1968}. For the 
system considered in this work, this approximation does not have 
significant effect on the results.

Equation \ref{eq:frequency_fit_general} then becomes

\begin{equation}\label{eq:frequency_fit}
\vec{\phi_c} = - \Omega^2 \vec{\upsilon_c}
\end{equation}

where $\vec{\phi_c} = Q^T M^{-\frac{1}{2}}  \vec{f_c}$ is the normal 
mode force and $\vec{\upsilon_c} = Q^T M^{\frac{1}{2}}\vec{u_c}$ the 
normal mode displacement. The $\vec{\theta}$ dependence is removed 
since we are interested in $\Omega$, which can be obtained from a
simple linear least squares fit, and how its elements are related 
to the parameters becomes irrelevant.

\begin{figure}
    \includegraphics[width=\columnwidth]{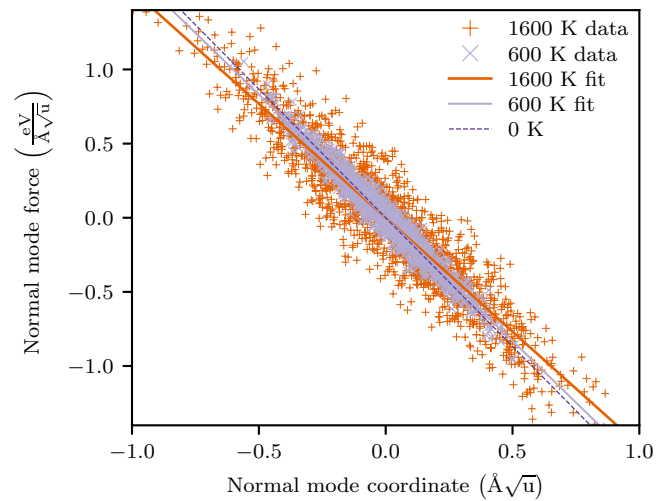}
    \caption{\label{fig:cloud_fits} A typical force-displacement plot 
    for a single frequency but symmetrically equivalent phonon modes 
    for illustrating and fitting effective harmonic models. 
    Not all points are shown for visualization purposes. As
    the temperature increases the distribution of forces for
    a fixed displacements broadens. The frequency of the mode
    is 20.6, 20.1 and 19.4~THz at 0~K, 600~K and 1600~K respectively.
    }
\end{figure}

For a perfectly harmonic crystal there exists one and only one 
$\Omega^2$. For anharmonic crystals however the uniqueness is lost due 
to the higher order terms in the Hamiltonian becoming relevant. A 
typical distribution of the normal mode force-displacement relation for 
a single phonon mode is depicted in Figures \ref{fig:cloud_histogram} 
and \ref{fig:cloud_fits}. The shape is that of a Gaussian due to both 
the nature of constant temperature sampling of the configurations and 
the anharmonicity of the system. The modes that are symmetrically 
equivalent can be placed on the same plot thereby increasing the amount 
of data, thereby reducing the uncertainty of the linear fit and 
retaining the symmetry properties of the vibrations.

As we have shown, TDEP is mathematically equivalent to SCAILD for 
systems with temperature independent phonon eigenvectors. The linear 
least squares fit is done in real space for the former and in normal 
mode space for the latter. 

Our approach differs slightly from the 
original formulation of SCAILD where sampling of the normal mode 
displacements, in order to obtain the forces, was limited to certain 
discrete values that give the same mean squared displacement as a quantum 
harmonic oscillator at a constant temperature, \ie

\begin{equation}\label{eq:displacements}
\vec{u} = M^{-\frac{1}{2}} Q \vec{d}
\end{equation}

where

\begin{equation}
d_i = \pm \sqrt{\frac{\hbar}{\Omega_{i,i}} \left[\frac{1}{2} + \left( \exp{\frac{\hbar \Omega_{i,i}}{k_B T}} -1 \right)^{-1} \right]}
\end{equation}

That makes it possible to take the average of $\Omega^2$ over all the sampled 
configurations which means the average slope of all the lines from the 
origin to every point in the force-displacement plot. This is not 
mathematically equivalent to a linear least squares fit through all the 
points. In addition, when the displacements for SCAILD are sampled from 
a Gaussian distribution, as done in this work, the most likely 
displacement for a given mode is zero while the force due to the 
anharmonicity is nonzero which leads to many large positive and 
negative $\Omega^2$ values which are not likely to cancel out leading 
to a large error in the estimation of the average value.

\subsection{Sampling of the configuration space}

The major difference between TDEP and SCAILD is the way the 
displacements are obtained. For TDEP it is a constant temperature 
molecular dynamics simulation consistent with the Hamiltonian of the 
anharmonic system whereas for SCAILD the displacements are sampled from 
the distribution self-consistent with the current best fitted harmonic 
Hamiltonian. The major advantage of the latter is the reduction in the 
number of calculations due to more efficient generation of uncorrelated 
samples as there exists an analytical expression that gives correct 
constant temperature statistics.

Using the samples generated by a canonical molecular dynamics or Monte 
Carlo simulation can be interpreted as finding the harmonic potential 
that reproduces the forces given by the anharmonic Hamiltonian for 
constant temperature with the smallest error in least square sense. It 
is not unlike using the method of force matching quite commonly 
employed in order to fit empirical interatomic potentials to DFT 
reference data \cite{Ercolessi1994}. Lack of transferability is often 
an issue with empirical potentials and, as we explain below, it also 
plays an important role in the effective harmonic models.

It is slightly less obvious what the sampling used in SCAILD method 
yields. In addition to the resulting model being harmonic, the 
displacements themselves are generated assuming non-interacting phonon 
modes. It has been suggested that this results in probing the wrong 
parts of the phase space \cite{Hellman2012} which is certainly true 
when the actual anharmonic system is concerned, however the goal here 
is not to perfectly reproduce the whole anharmonic Hamiltonian, but to 
pack the anharmonic vibrational information into a truncated Taylor 
expansion of the potential. Nevertheless by careful reasoning one can 
identify the possible effects of the harmonic sampling when compared to 
the anharmonic one. Since in the former case all the correlations in 
the motion of the normal modes are ignored, the probability of sampling 
higher energy configurations increases which results in larger magnitudes
for the sampled forces and in turn higher effective vibrational
frequencies which increase with temperature.

While it is clear that the two sampling methods are different, it does 
not automatically follow that one provides a better free energy 
estimation than the other. In either case we obtain a harmonic 
Hamiltonian both of which are just approximations and so are the 
predicted free energies. While there might exist an effective harmonic 
model that gives the exact free energy for the system, it is not 
immediately clear whether either of the methods should do that.

\subsection{Thermodynamic integration and free energy perturbation}

Since we expect SCAILD and TDEP to give different harmonic force 
constant matrices and therefore also different free energies, the full 
vibrational free energy needs to be calculated for a proper comparison. 
Both methods provide an excellent reference system for free energy 
perturbation (FEP) and thermodynamic integration (TI) when compared to 
the Einstein crystal \cite{DeKoning1996,Freitas2016} which has smaller 
overlap with the phase space of the anharmonic system, or a 0 K 
harmonic model \cite{Ryu2008a,Grabowski2009} which has undefined free 
energy in case of phonon modes with imaginary frequencies.

Both FEP and TI allow to calculate free energy differences from 
ensemble averages of energy differences. There are several variations 
of each. In this work we used a linear path through a parameter 
$\lambda$ from the effective harmonic system to the anharmonic one, in 
our case DFT system as follows:

\begin{equation}\label{eq:ulambda}
U(\lambda) = \lambda U_{\textrm{\tiny{DFT}}} + (1-\lambda) U_{\textrm{\tiny{EH}}}
\end{equation}

To get the free energy difference from thermodynamic integration
we calculate

\begin{equation}\label{eq:ti}
\Delta F = \int_0^1 \left\langle\frac{\partial U(\lambda)}{\partial\lambda}\right\rangle_{\lambda} d\lambda =
 \int_0^1 \left\langle U_{\textrm{\tiny{DFT}}} - U_{\textrm{\tiny{EH}}} \right\rangle_{\lambda} d\lambda
\end{equation}

In practice there are possibilities either to sample $\lambda$ on a 
discrete, but not necessarily equidistant grid or perform adiabatic 
switching where $\lambda$ changes continuously throughout the 
simulation.

Since SCAILD and TDEP sample a lot of configurations and provide both 
the harmonic and DFT energies, it is also in principle possible to 
calculate the anharmonic free energy without any extra steps through 
free energy perturbation.

\begin{equation}\label{eq:fep}
\Delta F = -1/\beta \cdot \ln \left \langle 
    \exp \left [ - \beta (U_B - U_A)\right ] \right \rangle _A
\end{equation}

For TDEP, $U_A$ are the potential energies from a DFT-MD calculation 
and $U_B$ the potential energies calculated using the fitted harmonic 
model. For SCAILD, $U_A$ are the harmonic potential energies calculated 
using the converged fit and $U_B$ the potential energies for the same 
structures obtained from DFT. FEP can also be done using stratified 
calculations. Similarly to thermodynamic integration several 
intermediate simulations using mixed Hamiltonians of the two endpoints 
can be done and the free energy differences between each step 
calculated using Equation \ref{eq:fep} and subsequently summed together 
\cite{Pohorille2010}.

\subsection{Calculations}
\begin{figure}
    \includegraphics[width=\columnwidth]{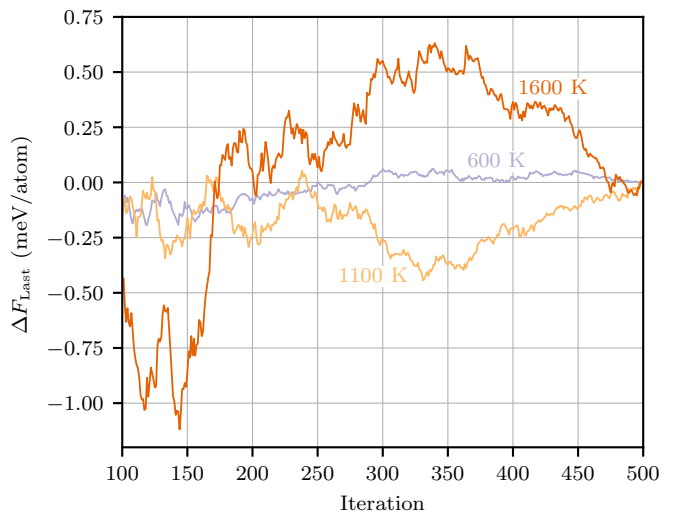}
    \caption{\label{fig:scaild_convergence} Convergence of the free 
    energy with respect to the number of iterations using SCAILD. The values are 
    given as a difference to the last step. The free energy difference
    between two consecutive iterations is typically orders of magnitude
    smaller than the longer term changes.}
\end{figure}

Cubic zirconia was chosen for the system to be studied for several 
reasons. It has been studied both using 0 K \cite{Parlinski1997}, self 
consistent lattice dynamics \cite{Souvatzis2008a} and other methods 
\cite{Sternik2005}. At ambient pressure it is not dynamically stable as 
indicated by large amount of imaginary phonon modes which disappear 
when a SCAILD analysis is performed. Another option for achieving 
dynamical stability for cubic ZrO2 is to put it under high pressure 
which we opted to do in order to retain the ability to compute the free 
energy from 0 K phonon calculation for comparison.

As explained above, the theory simplifies a lot in cases where the 
eigenvectors can be assumed not to depend on temperature. ZrO2 contains 
two types of atoms with relatively large difference in atomic weights 
which should theoretically allow for the possibility of temperature 
dependent eigenvectors. The size of the supercell was chosen to be $2 
\times 2 \times 2$ of the conventional unit cell with lattice parameter 
of $9.7$ Å containing a total of 96 atoms. It is not guaranteed to give 
a converged free energy with respect to the size, but is a good balance 
between capturing the overall vibrational, both harmonic and 
anharmonic, behavior while allowing for sufficiently long molecular 
dynamics simulations.

All phonon calculations were done using phonopy \cite{Togo2015} without 
applying non-analytical term correction. For SCAILD and TDEP our own 
implementations were used. For SCAILD we acquired the eigenvectors from 
phonopy after which the displacements were generated for every normal mode 
according to a Gaussian distribution with the same mean square 
displacement as given by theory for a classical harmonic oscillator at 
a constant temperature, \ie\ in Equation~\ref{eq:displacements}, $\vec{d}$
is obtained from

\begin{equation}
\sigma_i^2 = \left\langle d_i^2  \right\rangle = \frac{k_B T}{\Omega_{i,i}^2}
\end{equation}

The sampling was from a classical distribution in order to be directly 
comparable to classical DFT molecular dynamics, although at the 
investigated temperatures we expect the effect on the results to be 
relatively small.

As shown in Figure~\ref{fig:scaild_convergence} 
the changes in free energy between consecutive iterations can be orders 
of magnitude smaller than the changes over many iterations so instead 
of basing the convergence on the former we ran the simulations for a 
fixed number of (500) steps.

All the molecular dynamics simulations for TDEP were performed using 
Nosé-Hoover thermostatting with the Nosé mass corresponding to 
approximately a time period of 80~fs and velocity Verlet integration 
with a timestep of 1~fs. The symmetry requirements for the force 
constant matrix and crystal were taken into account when performing the 
linear fit of the displacement-force data. No additional cutoff 
distance in addition to that determined by the supercell was used for 
the force constants.

For both methods the simulations were done at 600, 1100 and 1600~K. At 
higher temperatures the oxygen atoms started migrating and occasionally 
creating vacancy-interstitial pairs which complicates the analysis not 
only for SCAILD and TDEP, but also for the following thermodynamic 
integration which set the upper limit used in this work. The number of 
MD steps was 20000 at 600 and 1100~K, and 30000 at 1600~K.

Thermodynamic integration was done for $\lambda$ values of 0, 0.1, 
0.25, 0.5, 0.75, 0.9 and 1. For TDEP $\lambda=1$ data is already 
acquired before the fitting and no extra calculation is needed. The 
same data was also used for SCAILD at $\lambda=1$. For SCAILD at 
$\lambda=0$ a separate fast MD run was done using the harmonic 
Hamiltonian after which a subset of uncorrelated samples was chosen for 
which the energy was calculated using DFT. Whereas Nosé-Hoover 
thermostat was used for TDEP mostly because it automatically retains 
the position of the center of mass of the system and therefore 
simplifies the fitting, Langevin dynamics was used for the 
thermodynamic integration steps in order to avoid the possibility of 
incorrect sampling of the phase space when the system is close to 
harmonic \cite{Martyna1992}. The drift of the center of mass was not 
problematic for TI since only the energy differences were obtained 
instead of displacements.

All DFT calculations were done using VASP 
\cite{Kresse1993,Kresse1994a,Kresse1996,Kresse1996a} with 
projector-augmented wave method \cite{Kresse1999} with a plane wave 
cutoff energy of 400 eV and GGA-PBE \cite{Perdew1996} XC-functional, 
\cite{Perdew1997}. The number of electrons treated explicitly was 12 
and 6 for Zr and O respectively. A  $2 \times 2 \times 2$ 
Monkhorst-Pack $\Gamma$-centered k-point grid \cite{Monkhorst1976a} and 
Gaussian smearing with $\sigma=0.05$ were used.

The choice for the k-point grid was based on the balance between 
sufficient accuracy and computational time. Fits for SCAILD and TDEP 
based on shorter runs with $3 \times 3 \times 3$ $\Gamma$-centered grid 
at 1600~K resulted in roughly 1~meV/atom difference. For thermodynamic 
integration a check was done by recalculating the $\lambda=0.5$ point 
for TDEP at 1600~K using a $4 \times 4 \times 4$ gamma-centered grid. 
The difference seen was within the statistical uncertainty of the 
original calculation so no extra up-sampling as done using the UP-TILD 
\cite{Grabowski2009} method was performed. 

As shown by Hellman \cite{Hellman2012}, the relative errors in forces 
decrease by orders of magnitude as the magnitude of the forces 
increases to that of the thermally excited states. Based on that the 0 
K harmonic calculations including obtaining the eigenvectors for SCAILD 
were done using a denser $7 \times 7 \times 7$ k-point grid.

\section{Results}

\begin{figure}
    \includegraphics[width=\columnwidth]{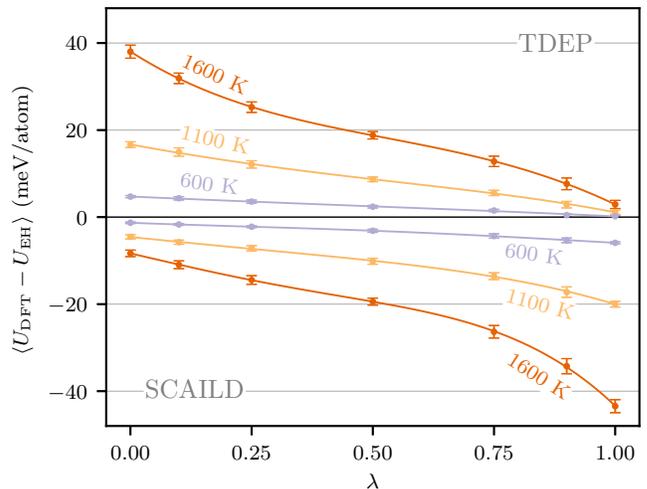}
    \caption{\label{fig:ti_lambda} Thermodynamic integration results 
    for TDEP and SCAILD. Every dot corresponds to one equilibrium
    MD calculation with a $\lambda$-mixed Hamiltonian. The error
    bars were obtained using block averaging method and are shown
    for $\pm 3$ standard deviations. The continuous line is a
    4th order polynomial fit with the reciprocals of the standard
    deviations used as weights.}
\end{figure}

\begin{table}
\caption{\label{tab:free_energies} Vibrational free energies (in 
meV/atom) calculated by different methods. The meaning of the acronyms 
is as follows: LD lattice dynamics, H harmonic, EH effective harmonic, 
AH anharmonic, FEP free energy perturbation, SFEP stratified free 
energy perturbation, TI thermodynamic integration. For TDEP the EH 
results are presented both for assuming temperature-independent and 
temperature-dependent eigenvectors denoted by $Q_0$ and $Q_T$ 
respectively.}

\begin{ruledtabular}
\bgroup
\def\arraystretch{1.4}
\begin{tabular}{llrrr}
Method  & Type & 600 K  & 1100 K & 1600 K\\
\hline
LD  & H &  4.0 & -176.7 & -415.9 \\
\hline
\multirow{4}{*}{SCAILD} & EH & 7.1 & -167.3 & -398.0 \\
 & AH,FEP & 4.1 & -176.4 & -415.1 \\
 & AH,SFEP & 3.8 & -177.9 & -419.0 \\
 & AH,TI & 3.8 & -178.1 & -419.1 \\ \hline
\multirow{5}{*}{TDEP} & EH,$Q_0$ & 0.9 & -186.8 & -438.5 \\
 & EH,$Q_T$ & 1.0 & -186.6 & -437.9  \\
 & AH,FEP & 3.2 & -180.1 & -422.0 \\
 & AH,SFEP & 3.4 & -177.8 & -418.4 \\
 & AH,TI & 3.5 & -177.7 & -418.6 \\
\end{tabular}
\egroup
\end{ruledtabular}
\end{table}

\begin{figure}
    \includegraphics[width=\columnwidth]{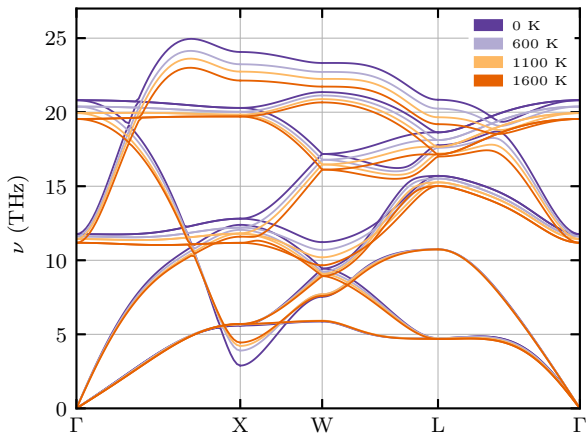} 
    \caption{\label{fig:phonon_tdep} Phonon dispersion dependence on 
    temperature as predicted by TDEP.}
\end{figure}

\begin{figure}
    \includegraphics[width=\columnwidth]{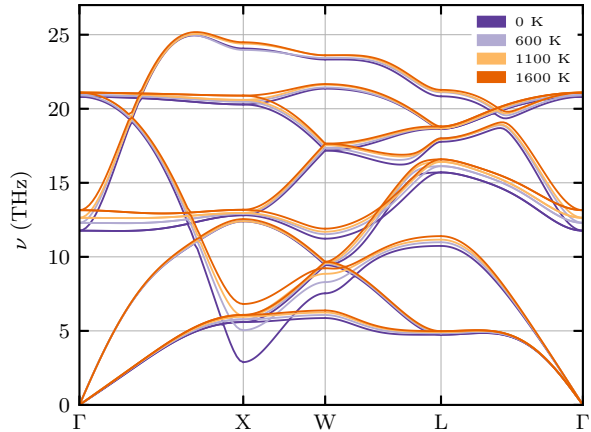}
    \caption{\label{fig:phonon_scaild} Phonon dispersion dependence on 
    temperature as predicted by SCAILD.}
\end{figure}

The vibrational free energies calculated using all of the methods are 
shown in Table~\ref{tab:free_energies}. The full anharmonic vibrational 
free energies are in good agreement. Even at the highest considered 
temperature, 1600~K, all stratified free energy perturbation and 
thermodynamic integration results are within 1~meV/atom. Both SCAILD 
and TDEP are however off by about 20 meV/atom and, perhaps 
surprisingly, even the free energy calculated from harmonic lattice 
dynamics is a lot closer to the value obtained from thermodynamic 
integration than those of the either effective harmonic method.

The error in free energies predicted by SCAILD and TDEP is opposite in 
sign. This is also evident from the thermodynamic integration results 
shown in Figure~\ref{fig:ti_lambda}. Since TDEP force constant matrix 
is fitted to the DFT-MD results also the average potential energy 
difference between DFT and TDEP is minimal at $\lambda=1$. It is also 
seen that setting the average potential energies to be equal as done in 
Ref~\citenum{Hellman2013a} (Equation 21) would do little to reduce the 
error. In this case the whole TI curve would be shifted down by roughly 
3 meV/atom for the temperature of 1600 K which accounts for only 15 \% 
of the total error. Similar conclusions can be drawn for SCAILD for 
which the agreement with DFT is best at $\lambda=0$ with the error due 
to mismatched average potential energies being slightly larger and if 
taken into account provides a better estimation for the full 
vibrational free energy than TDEP.

Free energy perturbation was not able to correct all of the discrepancy 
possibly due to the limited number of samples, however the stratified 
version gives free energies that agree very well with TI results. This 
is expected since the same input data was used for both methods.

In either case shifting the potential energy does not result in change 
in the force constants and therefore the phonon dispersions which are 
shown in Figures \ref{fig:phonon_tdep} for TDEP and 
\ref{fig:phonon_scaild} for SCAILD. The general trend is for TDEP to 
predict a decrease and for SCAILD an increase in frequencies over all 
of the reciprocal space. A notable exception to that is the lowest 
frequency mode at X-point which would be unstable without the external 
pressure. Both methods predict an increase in the frequency albeit 
SCAILD from 2.9~THz at 0~K to 6.1~THz, and TDEP to 4.4~THz at 1600~K.
We must note that since the calculations were done at fixed volume,
the typical lowering of the frequencies due to thermal expansion
is not taken into account.

The reason why TDEP, at least when up to second order force constants 
are considered, cannot provide a better estimate than SCAILD is that an 
effective harmonic Hamiltonian fitted at either $\lambda=0$ or at 
$\lambda=1$ exhibits similar non-transferability. Whereas the average 
energy difference between DFT and EH model can be forced to be zero at one of 
the endpoints of the TI curve, its absolute value can only 
monotonically increase as TI is performed and the sampled atomic 
configurations will start to differ from those used for the fit. This 
can be proven using Gibbs-Bogoliubov inequality \cite{Frenkel}, which 
states that

\begin{equation}
\left(\frac{\partial^2F}{\partial\lambda^2}\right)_{NVT} \le 0
\end{equation}

As the anharmonicity increases, the transferability and, thus, the 
estimate for the full vibrational free energy becomes worse. This can 
also be explained as follows: let us consider a full infinite expansion 
of the DFT Hamiltonian. Starting from a harmonic model we carry out the 
TDEP process, that is perform molecular dynamics runs at $\lambda=1$, 
and for each subsequent run increase the number of terms in the 
expansion. For up to quadratic term the effective harmonic model can be 
fitted exactly and $\Delta U = 0$ for any $\lambda$. As we add more 
terms $\Delta U$ at $\lambda = 0$ starts to deviate from $0$ more and 
more while being kept $0$ at $\lambda = 1$. Whereas LD ignores higher 
than second order terms, both SCAILD and TDEP incorporate those in an 
effective manner into the harmonic Hamiltonian and as shown, this can 
lead to a much larger error. In addition, it is not obvious that adding 
for example the third order force constants to the effective model 
Hamiltionan necessarily reduces the error, since the second order force 
constants may remain relatively unaffected.

\section{Conclusions}
Given temperature independent eigenvectors, the effective harmonic 
models obtained either through SCAILD or TDEP produce two types of 
systematic errors. One is due to the different choice of the ensemble 
of displacements and forces used to fit the data. Based on our results 
neither could be considered better than the other, the reason being 
that in either case it is not the remaining explicitly anharmonic 
energy obtained through thermodynamic integration that is minimized, 
but the potential energy difference at one of the endpoints of the 
$\lambda$-integration curve. The other systematic error, common to 
both, is introduced by trying to reproduce the anharmonic interactions 
in an effective harmonic manner.

The free energies predicted by either method diverge as the temperature 
increases with the general trend being SCAILD predicting an increase in 
the phonon frequencies and higher free energy than the exact value as 
opposed to TDEP which predicts decreasing frequencies and lower free 
energy. For the system considered in this work both give worse estimate 
of the free energy than 0 K harmonic model. Therefore relying on one or 
another can introduce large errors when investigating the vibrational 
thermodynamic properties of materials, especially when the results of 
either are compared with those obtained through other methods, for 
example, when using the quasiharmonic approximation to calculate the 
free energy of one phase, while one of the effective harmonic methods 
for another, in order to compute the phase stability. In addition, due 
to the non-linearity of the systematic error with respect to 
temperature, it is possible that the error in other thermodynamic 
quantities derived from free energy, such as heat capacity, will have 
even larger errors. Free energy perturbation can be used without extra 
computational cost in order to estimate the error, but since it may not 
account for all of the difference between the approximate and true free 
energy, thermodynamic integration or any other method that takes all of 
the anharmonicity explicitly into account has to be used.

\section{Acknowledgments}
The research leading to these results has received funding from the 
Swedish Centre for Nuclear Technology (SKC). Computational resources 
were provided by Swedish National Infrastructure for Computing (SNIC).
Comments by A.\ Samanta and A.\ Tamm from LLNL are greatly appreciated.

\bibliography{scaild_vs_tdep,vasp_cite}{}

\end{document}